\documentstyle[12pt]{article}
\begin{document}

\begin{titlepage}

\begin{flushright}
EPHOU-99-016 \\ 
November 1999 \ 
\end{flushright}

\vspace{20mm}

\begin{center}
{\large Supersymmetric Action of $SL(2;\mbox{\bf Z})$-covariant D3-brane \\
and its $\kappa$-symmetry}
\end{center}

\vspace{10mm}

\renewcommand{\thefootnote}{\fnsymbol{footnote}}

\begin{center}
Toshiya Suzuki \footnote[1]{tsuzuki@particle.sci.hokudai.ac.jp} 
\end{center}

\begin{center}
{\it Department of Physics, Hokkaido University \\
     Sapporo 060-0810, Japan } 
\end{center}

\vspace{15mm}

\begin{abstract}

We consider a supersymmetric extension of the $SL(2;\mbox{\bf Z})$-covariant D3-brane action
 proposed by Nurmagambetov, and prove its $\kappa$-symmetry in an on-shell 
 type-IIB supergravity background.

\end{abstract}

\begin{flushleft}

\vspace{25mm}


{\bf keywords} : D-brane, $\kappa$-symmetry, $SL(2;\mbox{\bf Z})$ duality 

\end{flushleft}

\end{titlepage}

\setcounter{footnote}{0}

The type-IIB string theory has the $SL(2;\mbox{\bf Z})$ duality \cite{HT}. Various branes
entering in the theory belong to $SL(2;\mbox{\bf Z})$ multiplets. Among
various type-IIB branes D3-brane is something special. It is an
$SL(2;\mbox{\bf Z})$ singlet and carries the charge of the background 4-form with
self-dual field strength. 
Furthermore it may also play a fundamental role in F theory. F theory is
a conjectural theory in $12$ dimensions, which explains the $SL(2;\mbox{\bf Z})$
duality from a geometric viewpoint \cite{V}. 
An idea that F theory is a theory of 3-brane propagating in $12$
dimensional spacetime has been proposed \cite{Ts,GG,JR}. If such an idea
is valid, the theory of D3-brane with a manifest
$SL(2;\mbox{\bf Z})$-covariance may be a crucial step toward the
construction of the theory of ``F3-brane''.  

$SL(2;\mbox{\bf Z})$-covariant type-IIB $p$-branes have been
considered by various authors \cite{To,CT,CW,Nu,WW}. In the cases of $p>1$, an important problem
 is the number of degrees of freedom. D-brane's collective coordinates
 include, in addition to the embedding fields into the target space, a
 vector field propagating in the 
 worldvolume. To construct an $SL(2;\mbox{\bf Z})$-covariant theory one needs to
 introduce $SL(2;\mbox{\bf Z})$ doublet worldvolume vector fields, but as a
 consequence of  
 this procedure one ends up with too many physical degrees of freedom. To resolve
 this problem, these 2 vector fields must not be independent; they
 should be related to each other under the electro-magnetic duality. One of the
 methods which 
 realize the electro-magnetic duality is the so-called PST
 formalism \cite{PST1,PST2,PST3}. Starting from the bosonic sector of
 the PST type M5-brane action \cite{PST4,BLNPST}, Nurmagambetov
 constructed an $SL(2;\mbox{\bf Z})$-covariant D3-brane action
 \cite{Nu}. \footnote{Also in \cite{Be1,Be2}, the relation between the
 D3-brane action and the M5-brane one is investigated with respect to the
 $SL(2;\mbox{\bf Z})$ duality.}

Nurmagambetov's model, however, includes only bosonic variables. One of
the most important properties of D-branes is that they are BPS-saturated 
objects, that is, they keep a half of the target space supersymmetries. So
one hopes to extend this model to supersymmetric branes. In general, the
theory of a supersymmetric extended object must have a local fermionic 
symmetry known as $\kappa$-symmetry
\cite{dAL1,dAL2,Si,GS,GHMNT,HLP,BST,AETW,APS1,APS2,CGNW,CGNSW,BT,BLNPST,APPS}, 
which halves the number of the fermionic degrees of
freedom and equates it with that of the bosonic ones. The purpose of this
letter is to prove the $\kappa$-symmetry of the supersymmetric version
of Nurmagambetov's $SL(2;\mbox{\bf Z})$-covariant D3-brane action. 

To start with, we review the $SL(2;\mbox{\bf Z})$-covariant D3-brane action,
proposed in \cite{Nu}. Its dynamical variables are the embedding
fields $x^{m}$ and the $SL(2;\mbox{\bf Z})$ doublet vector fields $A_{i , r}$ ($r=1,2$),
the field strength of which is defined as $F_{i j , r}=\partial_{i} A_{j 
, r} - \partial_{j} A_{i , r}$. The action is composed of 3 parts  
\begin{equation}
S = S_{DBI} + S_{PST} + S_{WZ} .
\label{action}
\end{equation}
The first part is a Dirac-Born-Infeld term
\begin{equation}
S_{DBI} = \int d^{4} \xi - \sqrt{-g} f \ , \ f = \sqrt{1 + {\tilde H}_{i , r} (\epsilon^{T} M \epsilon)^{r s} {{\tilde H}^{i}}_{, s} + \frac{1}{2} ({\tilde H}_{i , r} \epsilon^{r s} {\tilde H}_{j , s}) ({{\tilde H}^{i}}_{, t} \epsilon^{t u} {{\tilde H}^{j}}_{, u})} .
\label{bDBI}
\end{equation}
${\tilde H}_{i , r}$ is defined as 
\begin{equation}
{\tilde H}_{i , r} = v^{j} *H_{j i , r} \ , {*H^{i j}}_{, r} = \frac{\epsilon^{i j k l}}{2! \sqrt{-g}} H_{k l , r} ,
\label{tilde}
\end{equation}
where $H_{i j , r}$ is the improved field strength $H_{i j , r}=F_{i j
,r}-C_{i j , r}$ ($C_{i j , r}$ is the background 2-form) and $v^{i}$ is 
defined in terms of the worldvolume scalar $a$ as
\begin{equation}
v^{i} = \frac{\partial^{i} a}{\sqrt{-(\partial a)^{2}}} .
\label{v}
\end{equation}
Notice that $v^{i} {\tilde H}_{i , r} = 0$ and $(v)^{2}=-1$. $M$ is
written in terms of the background scalars, dilaton $\phi$ and axion
$\chi$
\begin{equation}
M = \frac{1}{e^{-\phi}}
\left[
\begin{array}{cc}
1 & \chi \\
\chi & \chi^{2} + e^{-2 \phi}
\end{array}
\right] .
\label{M}
\end{equation}
The second part is a Pasti-Sorokin-Tonin term
\begin{equation}
S_{PST} = \int d^{4} \xi \sqrt{-g} \frac{1}{2} {\tilde H}_{i , r} \epsilon^{r s} {H^{i j}}_{, s} v_{j} ,
\label{bPST}
\end{equation}
and the third part a Wess-Zumino term
\begin{equation}
S_{WZ} = \int C^{(4)} - \frac{1}{2} F^{(2)}_{r} C^{(2)}_{s} \epsilon^{r s} ,
\label{bWZ}
\end{equation}
where $C^{(4)}$ is the background 4-form.

This action has the following local symmetries; \\
\underline{one usual gauge symmetry}
\begin{equation}
\delta A_{i , r} = \partial_{i} \phi_{r} , 
\label{sym1}
\end{equation}
and \underline{two characteristic symmetries}
\begin{equation}
\delta A_{i , r} = \partial_{i} a \cdot  \varphi_{r} \ , \ \delta a = 0 ,
\label{sym2}
\end{equation}
and
\begin{equation}
\delta a = \Phi \ , \ \delta A_{i , r} = \frac{\Phi}{\sqrt{-(\partial a)^{2}}} \epsilon_{r s} {{\cal G}_{i}}^{, s} ,
\label{sym3}
\end{equation}
where
\begin{equation}
{{\cal G}_{i}}^{, r} = {{\cal V}_{i}}^{, r} - \epsilon^{r s} H_{i j , s} v^{j} \ , \ {{\cal V}_{i}}^{, r} = \frac{\delta f}{\delta {{\tilde H}^{i}}_{, r}} .
\label{calGV}
\end{equation}
To see these symmetries, consider the variation $\delta S$ under $\delta A_{i , r}$
and  $\delta a$
\begin{equation}
\delta S = - \int d^{4} \xi \epsilon^{i j k l} \{ \delta A_{i ,r} - \frac{\delta a}{\sqrt{-(\partial a)^{2}}} \epsilon_{r s} {{\cal G}_{i}}^{, s} \} \partial_{j} [v_{k} {{\cal G}_{l}}^{, r}] .
\label{d123}
\end{equation}
The symmetry (\ref{sym2}) was used to reduce the equation of motion for 
$A_{i , r}$ to the non-linear duality equation ${{\cal G}_{i}}^{, r}=0$ \cite{PST1,PST2,PST3,Nu}, 
which is necessary for keeping the number of the physical degrees of
freedom of the D3-brane to be 8.
On the other hand, the symmetry (\ref{sym3}) can be used to remove the
auxiliary field $a$. Fixing this symmetry by a condition $\partial_{i} a 
= \delta_{i}^{0}$, one obtains an action which is no longer manifestly
worldvolume general coordinate covariant. Solving one of the
$SL(2;\mbox{\bf Z})$ doublet field strength in terms of the other, one
recovers the conventional D3-brane action \cite{Be1,KP}.

Now, we rewrite the action in a manifestly $SL(2;\mbox{\bf Z})$-covariant
manner. To this end we introduce the complex $SL(2;\mbox{\bf Z})$ doublet
background scalars $u^{r}$ \cite{SW,Sc,HW}, which satisfy an $SL(2;\mbox{\bf Z})$-invariant
constraint 
\begin{equation}
\frac{i}{2} \epsilon_{r s} u^{r} \bar{u}^{s} = 1 . 
\label{u}
\end{equation}
The physical background scalars, $\phi$ and $\chi$, belong to the
coset space $SL(2;\mbox{\bf R})/U(1)$ 
\begin{equation}
\tau = - \frac{u^{1}}{u^{2}} \ , \ \tau = \chi + i e^{- \phi} .
\label{tau}
\end{equation}
Using $u$'s, one can define various $SL(2;\mbox{\bf Z})$ invariants; \\
from $u$'s only one obtains
\begin{equation}
Q = \frac{1}{2} \epsilon_{r s} d u^{r} \bar{u}^{s} ,
\label{Q}
\end{equation}
and
\begin{equation}
P = \frac{1}{2} \epsilon_{r s} d u^{r} u^{s} \ , \ \bar{P} = \frac{1}{2} \epsilon_{r s} d \bar{u}^{r} \bar{u}^{s} .
\label{P}
\end{equation}
Together with $SL(2;\mbox{\bf Z})$ doublets,
\begin{equation}
{\cal C}^{(2)} = u^{r} C^{(2)}_{r} \ , \ \bar{\cal C}^{(2)} = \bar{u}^{r} C^{(2)}_{r} ,
\label{calC}
\end{equation}
and
\begin{equation}
{\cal H}^{(2)} = u^{r} H^{(2)}_{r} \ , \ \bar{\cal H}^{(2)} = \bar{u}^{r} H^{(2)}_{r} ,
\label{calH}
\end{equation}
etc.

In terms of these $SL(2;\mbox{\bf Z})$ invariants, the action is expressed as
\begin{equation}
S_{DBI} = \int d^{4} \xi - \sqrt{-g} f \ , \ f = \sqrt{1 + \bar{\tilde{\cal H}}_{i} \tilde{\cal H}^{i} + \frac{1}{2} Im(\bar{\tilde{\cal H}}_{i} \tilde{\cal H}_{j}) Im(\bar{\tilde{\cal H}}^{i} \tilde{\cal H}^{j})} ,
\label{sDBI}
\end{equation}
\begin{equation}
S_{PST} = \int d^{4} \xi \sqrt{-g} \frac{1}{2} Im(\bar{\tilde{\cal H}}_{i} {\cal H}^{i j}) v_{j} ,
\label{sPST}
\end{equation}
\begin{equation}
S_{WZ} = \int C^{(4)} - \frac{1}{2}Im(\bar{\cal F}^{(2)} {\cal C}^{(2)}) .
\label{sWZ}
\end{equation}
Fixing the $U(1)$ symmetry by a condition $Im(u^{2})=0$, one obtains 
(\ref{bDBI},\ref{bPST},\ref{bWZ}).

The supersymmetric action is identical to the bosonic one, but with the
target space replaced by the type-IIB superspace with coordinates
$z^{M}=(x^{m},\theta^{\mu},\theta^{\bar{\mu}})$, where $\theta^{\mu}$ is a 10
dimensional Weyl spinor and $\theta^{\bar{\mu}}$ is the complex conjugate 
of it. Like other supersymmetric branes, this action must
have the $\kappa$-symmetry. In the rest, we will prove this symmetry.

The $\kappa$ transformation is a kind of translation, which satisfies
\begin{eqnarray}
i_{\delta z} E^{a} &=& 0 \nonumber \\
i_{\delta z} E^{\alpha} = \kappa^{\alpha} &,& i_{\delta z} E^{\bar{\alpha}} = \kappa^{\bar{\alpha}} ,
\label{kappa}
\end{eqnarray}
where $i_{\delta z}$ is the symbol for inner product: the inner product
of a vector $V^{N}$ and a p-form $\Omega = \frac{1}{p!} d z^{M_{1}}
... d z^{M_{p}} \Omega_{M_{p} ... M_{1}}$ is defined as 
\begin{equation}
i_{V} \Omega = \frac{1}{(p-1)!} d z^{M_{1}} ... d z^{M_{p-1}} V^{N} \Omega_{N M_{p-1} ... M_{1}} .
\label{inner}
\end{equation}
$E^{A}$ is the basis 1-form of the local Lorentz frame, and is related 
to the coordinate 1-form $d z^{M}$
\begin{equation}
E^{A} = d z^{M} {E_{M}}^{A},
\end{equation}
where ${E_{M}}^{A}$ is the super vielbein. The torsion 2-form is defined
as the covariant derivative of $E^{A}$
\begin{equation}
T^{A} = D E^{A} = d E^{A} + E^{B} \hat{\omega}_{B}^{ \ \ A} ,
\end{equation}
where $\hat{\omega}_{A}^{ \ \ B}$ is the connection 1-form. The
variation of a background form $\Omega$ is given by its Lie derivative   
\begin{equation}
\delta \Omega = {\cal L}_{\delta z} \Omega = (i_{\delta z} d + d i_{\delta z}) \Omega .  
\label{Lie}
\end{equation}
Using this formula, one obtains $\delta Q$, $\delta P$, $\delta
C^{(4)}$, $\delta {\cal C}^{(2)}$ and $\delta E^{A}$, from which one can
read off the variation of 
the induced metric $g_{i j} = {E_{i}}^{a} {E_{j}}^{b} \eta_{a b}$ 
\begin{equation}
\delta g_{i j} = 2 {E_{( i}}^{a} {E_{j )}}^{B} \kappa^{\alpha} {T_{B \alpha}}^{b} \eta_{a b} + c.c. .
\label{dg}
\end{equation}
One should take $\kappa$, rather than $\delta z$, as the transformation
parameter. The transformation of $A_{i , r}$ should be
specified in such a way that $\delta H_{i j , r}$ is $\kappa$-covariant,
that is, contains no derivatives of $\kappa$. So one takes $\delta
A^{(1)}_{r} = i_{\delta z} C^{(2)}_{r}$, and then obtains
\begin{equation}
\delta {\cal H}_{i j} = -i \bar{\cal H}_{i j} (i_{\delta z} P) - {E_{j}}^{C} {E_{i}}^{B} (i_{\delta z} {\cal R}^{(3)})_{B C} + i {\cal H}_{i j} (i_{\delta z} Q) . 
\label{dH}
\end{equation}
${\cal R}^{(3)}$ is defined by the same manner as
(\ref{calC},\ref{calH})
\begin{equation}
{\cal R}^{(3)} = u^{r} R_{r}^{(3)} \ , \ \bar{\cal R}^{(3)} = \bar{u}^{r} R_{r}^{(3)} ,
\end{equation}
where $R_{r}^{(3)} = d C_{r}^{(2)}$. We also define the 5-form field
strength $R^{(5)}$ as
\begin{equation}
R^{(5)} = d C^{(4)} - \frac{1}{2} C_{r}^{(2)} R_{s}^{(3)} \epsilon^{r s} .
\end{equation}
In what follows we set $\delta a = 0$.

To proceed to the calculation of the $\kappa$ variation, we need the
knowledge of the components of the background fields. They are given by 
the supergravity constraint, which is equivalent to the equation of
motion of the supergravity. We use the constraint given in \cite{CW}
\footnote{The normalization of the background 4-form is different from
the one in \cite{CW} by a factor 2.} \\
(dimension $0$ components)
\begin{eqnarray}
{T_{\alpha \bar{\beta}}}^{a} & = & (\Gamma^{a})_{\alpha \beta} , \nonumber \\
{\cal R}_{a \alpha \beta} & = & 2 (\Gamma_{a})_{\alpha \beta} , \nonumber \\
R_{a b c \alpha \bar{\beta}} & = & i (\Gamma_{a b c})_{\alpha \beta} ,
\label{IIB0}
\end{eqnarray}
(dimension $\frac{1}{2}$ components)
\begin{eqnarray}
R_{a b c d \alpha} & = & 0 , \nonumber \\
{\cal R}_{a b \bar{\alpha}} & = & -i (\Gamma_{a b} P)_{\alpha} \nonumber , \\
{T_{\bar{\alpha} \bar{\beta}}}^{\gamma} & = & i {\delta_{( \alpha}}^{\gamma} P_{\beta )} - \frac{i}{2} (\Gamma_{a})_{\alpha \beta} (\Gamma^{a} P)^{\gamma} , \nonumber \\
P_{\bar{\alpha}} & = & 0 , \nonumber \\
Q_{\alpha} & = & 0 .
\label{IIB1/2}
\end{eqnarray}
The components of dimension $1$ and above do not appear in the $\kappa$
variation. Those relevant to the $\kappa$ variation which are not
contained in (\ref{IIB0},\ref{IIB1/2}) vanish. $\Gamma_{a}$ is the 10
dimensional $\Gamma$ matrix, and we use the real representation so that
we do not distinguish $\bar{\alpha}$ from $\alpha$ in the R.H.S. of
(\ref{IIB0},\ref{IIB1/2}). In the following, we use $\gamma_{i} =
{E_{i}}^{a} \Gamma_{a}$. We also define $\gamma^{(4)} = \gamma_{0}
\gamma_{1} \gamma_{2} \gamma_{3}$ and so $(\gamma^{(4)})^{2} = g$. We
should note that the complex conjugation of the product of fermionic
quantities is defined without a reverse of the order. 

Using the above constraint, we proceed to the calculation of the
$\kappa$ variation 
\begin{equation}
\delta {\cal L} = \delta {\cal L}^{[\frac{1}{2}]} + \delta {\cal L}^{[0]} .
\label{dL}
\end{equation}
Here we decompose $\delta {\cal L}$ into 2 parts, $\delta {\cal
L}^{[\frac{1}{2}]}$ which contains dimension $\frac{1}{2}$ component
$P_{\alpha}$ ($\bar{P}_{\bar{\alpha}}$), and $\delta {\cal L}^{[0]}$
which does not. 

$\delta {\cal L}^{[\frac{1}{2}]}$ is further decomposed into 2 parts, one 
from the DBI term and the other from the PST and WZ terms
\begin{equation}
\delta {\cal L}^{[\frac{1}{2}]} = \delta {\cal L}_{DBI}^{[\frac{1}{2}]} + \delta {\cal L}_{PST + WZ}^{[\frac{1}{2}]} .
\label{dL1/2}
\end{equation}
$\delta {\cal L}_{DBI}^{[\frac{1}{2}]}$ is given by
\begin{equation}
\delta {\cal L}_{DBI}^{[\frac{1}{2}]} = \frac{- \sqrt{-g}}{2 f} [W^{(1)} + W^{(2)} + W^{(3)}] ,
\label{dLW}
\end{equation}
where
\begin{equation}
W^{(1)} = - \frac{\epsilon^{i j k l}}{2! \sqrt{-g}} v_{i} (i \kappa \gamma_{j k} \bar{P} \tilde{\cal H}_{l} - i \bar{\kappa} \gamma_{j k} P \bar{\tilde{\cal H}}_{l}) ,
\label{W1} 
\end{equation}
\begin{equation}
W^{(2)} = i \bar{\kappa} \bar{P} ({\tilde{\cal H}}_{i} {\tilde{\cal H}}^{i}) - i \kappa P (\bar{{\tilde{\cal H}}}_{i} \bar{{\tilde{\cal H}}}^{i}) ,
\label{W2}
\end{equation}
\begin{eqnarray}
W^{(3)} & = & \frac{\epsilon^{i j k l}}{2 \cdot 2! \sqrt{-g}} v_{i} [i \kappa \gamma_{j k} \bar{P} \{ \bar{\tilde{\cal H}}_{l} (\tilde{\cal H}_{m} \tilde{\cal H}^{m}) - \tilde{\cal H}_{l} (\bar{{\tilde{\cal H}}}_{m} \tilde{\cal H}^{m}) \} \nonumber \\
 & & -i \bar{\kappa} \gamma_{j k} P \{ \tilde{\cal H}_{l} (\bar{{\tilde{\cal H}}}_{m} \bar{{\tilde{\cal H}}}^{m}) - \bar{\tilde{\cal H}}_{l} ({\tilde{\cal H}}_{m} \bar{{\tilde{\cal H}}}^{m}) \}] ,
\label{W3}
\end{eqnarray}
and $\delta {\cal L}^{[\frac{1}{2}]}_{PST+WZ}$ is 
\begin{equation}
\delta {\cal L}_{PST + WZ}^{[\frac{1}{2}]} = X^{(1)} ,
\label{dLX}
\end{equation}
where
\begin{equation}
X^{(1)} = - \frac{\sqrt{-g}}{2} v^{i} (\bar{\tilde{\cal H}}^{j} \bar{\kappa} \gamma_{i j} P + \tilde{\cal H}^{j} \kappa \gamma_{i j} \bar{P}) .
\label{X1}
\end{equation}
In (\ref{W1} - \ref{W3}) and (\ref{X1}), we use $\bar{\kappa}$ to denote
$\kappa^{\bar{\alpha}}$. 

$\delta {\cal L}^{[0]}$ is also decomposed into 2 parts
\begin{equation}
\delta {\cal L}^{[0]} = \delta {\cal L}_{DBI}^{[0]} + \delta {\cal L}_{PST + WZ}^{[0]} .
\label{dL0}
\end{equation}
Here $\delta {\cal L}_{DBI}^{[0]}$ is
\begin{equation}
\delta {\cal L}_{DBI}^{[0]} = \frac{- \sqrt{-g}}{2 f} [Y^{(0)} + Y^{(1)} + Y^{(2)} + Y^{(3)} + Y^{(4)}] ,
\label{dLY}
\end{equation}
where
\begin{equation}
Y^{(0)} = -2 (\kappa \gamma^{i} \bar{E}_{i} + \bar{\kappa} \gamma^{i} E_{i}) ,
\label{Y0}
\end{equation}
\begin{equation}
Y^{(1)} = - \frac{2}{\sqrt{-g}}(\kappa v^{i} \bar{\tilde{\cal H}}^{j} \gamma_{i j k} \gamma^{(4)} E^{k} + \bar{\kappa} v^{i} \tilde{\cal H}^{j} \gamma_{i j k} \gamma^{(4)} \bar{E}^{k} ) ,
\label{Y1}
\end{equation}
\begin{eqnarray}
Y^{(2)} & = & \kappa \bar{\tilde{\cal H}}^{i} \tilde{\cal H}^{j} \{ 2 (v^{k} \gamma_{k}) g_{i j} v^{l} - (\gamma_{i} \delta_{j}^{l} + \gamma_{j} \delta_{i}^{l}) \} \bar{E}_{l} \nonumber \\
 & & + \bar{\kappa} \tilde{\cal H}^{i} \bar{\tilde{\cal H}}^{j} \{ 2 (v^{k} \gamma_{k}) g_{i j} v^{l} - (\gamma_{i} \delta_{j}^{l} + \gamma_{j} \delta_{i}^{l}) \} E_{l} ,
\label{Y2}
\end{eqnarray}
\begin{eqnarray}
Y^{(3)} & = & - \frac{1}{\sqrt{-g}} [ \kappa \{ (\bar{\tilde{\cal H}}_{l} \tilde{\cal H}^{l}) v^{i} \bar{\tilde{\cal H}}^{j} \gamma_{i j k} \gamma^{(4)} - (\bar{\tilde{\cal H}}_{l} \bar{\tilde{\cal H}}^{l}) v^{i} \tilde{\cal H}^{j} \gamma_{i j k} \gamma^{(4)} \} E^{k} \nonumber \\
 & & + \bar{\kappa} \{ (\tilde{\cal H}_{l} \bar{\tilde{\cal H}}^{l}) v^{i} \tilde{\cal H}^{j} \gamma_{i j k} \gamma^{(4)} - (\tilde{\cal H}_{l} \tilde{\cal H}^{l}) v^{i} \bar{\tilde{\cal H}}^{j} \gamma_{i j k} \gamma^{(4)} \} \bar{E}^{k} ] ,
\label{Y3}
\end{eqnarray}
\begin{eqnarray}
Y^{(4)} & = & \frac{1}{2} [ \{ (\bar{\tilde{\cal H}}_{k} \tilde{\cal H}^{k})^{2} - (\tilde{\cal H}_{k} \tilde{\cal H}^{k}) (\bar{\tilde{\cal H}}_{l} \bar{\tilde{\cal H}}^{l}) \} (g^{i j} + 2 v^{i} v^{j}) \nonumber \\
 & & - (\bar{\tilde{\cal H}}_{k} \tilde{\cal H}^{k}) (\bar{\tilde{\cal H}}^{i} \tilde{\cal H}^{j} + \tilde{\cal H}^{i} \bar{\tilde{\cal H}}^{j}) + (\bar{\tilde{\cal H}}_{k} \bar{\tilde{\cal H}}^{k}) \tilde{\cal H}^{i} \tilde{\cal H}^{j} + (\tilde{\cal H}_{k} \tilde{\cal H}^{k}) \bar{\tilde{\cal H}}^{i} \bar{\tilde{\cal H}}^{j} ] \nonumber \\
 & & (\kappa \gamma_{i} \bar{E}_{j} + \bar{\kappa} \gamma_{i} E_{j}) ,
\label{Y4}  
\end{eqnarray}
and $\delta {\cal L}_{PST + WZ}^{[0]}$ is
\begin{equation}
\delta {\cal L}_{PST + WZ}^{[0]} = Z^{(0)} + Z^{(1)} + Z^{(2)} ,
\label{dLZ}
\end{equation}
where
\begin{equation}
Z^{(0)} = i \kappa \gamma^{(4)} \gamma^{i} \bar{E}_{i} - i \bar{\kappa} \gamma^{(4)} \gamma^{i} E_{i} ,
\label{Z0}
\end{equation}
\begin{equation}
Z^{(1)} = i \sqrt{-g} \{ (\bar{\tilde{\cal H}}^{i} v^{j} - \bar{\tilde{\cal H}}^{j} v^{i}) \kappa \gamma_{i} E_{j} - ( \tilde{\cal H}^{i} v^{j} - \tilde{\cal H}^{j} v^{i}) \bar{\kappa} \gamma_{i} \bar{E}_{j} \} ,
\label{Z1}
\end{equation}
\begin{eqnarray}
Z^{(2)} & = & \frac{i}{2} \kappa  \gamma^{(4)} [ \bar{\tilde{\cal H}}^{i} \tilde{\cal H}^{j} \gamma_{i j} \{ \gamma^{k} + 4 (v^{l} \gamma_{l}) v^{k} \} + \bar{\tilde{\cal H}}^{i} \tilde{\cal H}^{j} \{ ({\delta_{i}}^{k} \gamma_{j} - {\delta_{j}}^{k} \gamma_{i}) -2 v^{l} \gamma_{l i j} v^{k} \} ] \bar{E}_{k} \nonumber \\
 & & - \frac{i}{2} \bar{\kappa} \gamma^{(4)} [ \tilde{\cal H}^{i} \bar{\tilde{\cal H}}^{j} \gamma_{i j} \{ \gamma^{k} + 4 (v^{l} \gamma_{l}) v^{k} \} + \tilde{\cal H}^{i} \bar{\tilde{\cal H}}^{j} \{ ({\delta_{i}}^{k} \gamma_{j} - {\delta_{j}}^{k} \gamma_{i}) -2 v^{l} \gamma_{l i j} v^{k} \} ] E_{k} . \nonumber \\
 & & 
\label{Z2}
\end{eqnarray}
In (\ref{Y0} - \ref{Y4}) and (\ref{Z0} - \ref{Z2}), $\bar{E}_{i}$ stands 
for ${E_{i}}^{\bar{\alpha}}$. 

To complete the proof of the $\kappa$-symmetry, one must find a matrix $\Gamma$, which
satisfies $(\Gamma)^{2} = 1$ and $tr \Gamma = 0$. Consider a matrix
\begin{equation}
\Gamma = \frac{1}{\sqrt{-g} f} 
\left[
\begin{array}{cc}
i \gamma^{(4)} - \frac{ig}{4} \epsilon_{i j k l} \bar{\tilde{\cal H}}^{i} \tilde{\cal H}^{j} \gamma^{k l} &
i \sqrt{-g} \bar{\tilde{\cal H}}^{i} v^{j} \gamma_{i j} \\
-i \sqrt{-g} \tilde{\cal H}^{i} v^{j} \gamma_{i j} &
-i \gamma^{(4)} + \frac{ig}{4} \epsilon_{i j k l} \tilde{\cal H}^{i} \bar{\tilde{\cal H}}^{j} \gamma^{k l}
\end{array}
\right] .
\label{Gamma}
\end{equation}
One can check that $tr \Gamma = 0$, which is trivial, and $(\Gamma)^{2} =
1$. Using $\Gamma$, projection matrices ${\cal P}_{\pm}$ are defined as 
\begin{equation}
{\cal P}_{\pm} = \frac{1}{2} (1 \pm \Gamma) .
\label{proj}
\end{equation}
After some works, one can show that $\delta {\cal
L}^{[\frac{1}{2}]}$ and $\delta {\cal L}^{[0]}$ are combined into
\begin{equation}
\delta {\cal L}^{[\frac{1}{2}]} = [\kappa \ \bar{\kappa}] {\cal P}_{+} 
\left[
\begin{array}{c}
- \sqrt{-g} v^{i} \tilde{\cal H}^{j} \gamma_{i j} \bar{P} \\
- \sqrt{-g} v^{i} \bar{\tilde{\cal H}}^{j} \gamma_{i j} P
\end{array}
\right] 
\label{kappa1/2}
\end{equation}
and
\begin{eqnarray}
\delta {\cal L}^{[0]} & = & [\kappa \ \bar{\kappa}] {\cal P}_{+} \nonumber \\
 & & 
\left[
{\begin{array}{c}
2i \gamma^{(4)} \gamma^{i} \bar{E}_{i} + 2i \sqrt{-g} (\bar{\tilde{\cal H}}^{i} v^{j} - \bar{\tilde{\cal H}}^{j} v^{i}) \gamma_{i} E_{j} \\
+i \gamma^{(4)} [ \bar{\tilde{\cal H}}^{i} \tilde{\cal H}^{j} \gamma_{i j} \{ \gamma^{k} + 4 (v^{l} \gamma_{l}) v^{k} \} + \bar{\tilde{\cal H}}^{i} \tilde{\cal H}^{j} \{ ({\delta_{i}}^{k} \gamma_{j} - {\delta_{j}}^{k} \gamma_{i}) -2 v^{l} \gamma_{l i j} v^{k} \} ] \bar{E}_{k} \\
 \\
 \\
-2i \gamma^{(4)} \gamma^{i} E_{i} - 2i \sqrt{-g} (\tilde{\cal H}^{i} v^{j} - \tilde{\cal H}^{j} v^{i}) \gamma_{i} \bar{E}_{j} \\
-i \gamma^{(4)} [ \tilde{\cal H}^{i} \bar{\tilde{\cal H}}^{j} \gamma_{i j} \{ \gamma^{k} + 4 (v^{l} \gamma_{l}) v^{k} \} + \tilde{\cal H}^{i} \bar{\tilde{\cal H}}^{j} \{ ({\delta_{i}}^{k} \gamma_{j} - {\delta_{j}}^{k} \gamma_{i}) -2 v^{l} \gamma_{l i j} v^{k} \} ] E_{k} \\
\end{array}}
\right] , \nonumber \\
 & & 
\label{kappa0}
\end{eqnarray}
respectively. Clearly, when $[\kappa \ \bar{\kappa}] = [\zeta \
\bar{\zeta}] {\cal P}_{-}$ ($\zeta$ is an unconstrained spinor), $\delta
{\cal L}$ vanishes, {\it i.e.}, a half of fermions are not physical.

In this article, we proved the $\kappa$-symmetry of the 
$SL(2;\mbox{\bf Z})$-covariant super D3-brane action. Like other
supersymmetric brane cases, the background supergravity constraint
turned out to be a sufficient 
condition for the $\kappa$-symmetry. Though we have not shown that this
is also a necessary one, this model has passed a non-trivial test
of the consistency with the type-IIB supergravity \footnote{In
\cite{Nu}, Nurmagambetov discusses the coupling of the D3-brane to the
bosonic sector of the type-IIB supergravity constructed by the PST
formalism \cite{DLS,DLT}.} . 

Topics concerning the supersymmetry have been major points of interest in the
D-brane physics. In a future work, the supersymmetry algebra
\cite{ST,Ha,HK1,HK2}, BPS saturated worldvolume solitons 
\cite{CM,Gi,HLW1,HLW2} etc. need to be also investigated.

To explain the origin of the $SL(2;\mbox{\bf Z})$ duality of the type-IIB
theory has been a main motivation of F theory. In this respect it may be
interesting to examine whether the theory of $SL(2;\mbox{\bf Z})$-covariant
super D3-brane may have a 12 dimensional interpretation.  

\begin{flushleft}
{\large Acknowledgement}
\end{flushleft}

The author would like to thank Prof. R. Nakayama for careful reading of
the manuscript.

\end{document}